\documentclass[12pt]{iopart}
\usepackage[utf8x]{inputenc}
\usepackage{graphicx}
\usepackage{bm}
\usepackage{fancyhdr}
\begin{document}
\fancyhead{}
\fancyfoot{}
\lhead{T. Karin \emph{et al.}}
\rhead{\thepage}
\title{Transport of charged particles by adjusting rf voltage amplitudes}
\author{Todd Karin\(^1\)\footnote{Present address: Physics Department, University of Washington Seattle, US} , Isabela Le Bras\(^1\) \footnote{Present address: Deptartment of Earth, Atmospheric and Planetary Sciences, Massachusetts Institute of Technology, US}  \footnote{Both first authors contributed equally to this work.}, Andreas Kehlberger\(^2\), Kilian Singer\(^2\), Nikos Daniilidis\(^1\), and Hartmut H\"affner\(^1\)}
\address{1. Physics Department, University of California Berkeley, US}
\address{2. Institut f\"ur Physik, Johannes Gutenberg Universit\"at Mainz, 55128 Mainz, Germany}
\ead{tkarin@uw.edu, hhaeffner@berkeley.edu}
\begin{abstract}
We propose a planar architecture for scalable quantum information processing (QIP) that includes X-junctions through which particles can move without micromotion.
This is achieved by adjusting radio frequency (rf) amplitudes to move an rf null along the legs of the junction.
We provide a proof-of-principle by transporting dust particles in three dimensions via adjustable rf potentials in a 3D trap.
For the proposed planar architecture, we use regularization techniques to obtain amplitude settings that guarantee smooth transport through the X-junction.
\end{abstract}
\maketitle

\section{Introduction}

Trapped ions present a promising avenue for scalable quantum computing \cite{Kielpinski2002, Haffner08}. One of the outstanding questions in ion trap quantum computing is how to accomplish two-qubit operations between arbitrary ions in large quantum registers, thus enabling large-scale quantum information processing. Current efforts focus on transporting the ions close to one another \cite{Wineland98,Kielpinski2002}. Most encouragingly, the major building blocks of such a system have been shown to work \cite{Monroe95, Sackett00, Schmidt-Kaler03, Home09, Hanneke09}.

Junctions (places where ions can be sent in three or more directions) are
one way to bring arbitrary ions close to one another. Currently, three-dimensional trap structures allow high-fidelity transport of ions through three and four-way junctions while keeping motional heating low and preserving the coherence of the electronic degree-of-freedom \cite{Hensinger06,Blakestad09,Blakestad2011}. For scalable quantum information processing (QIP), planar traps have advantages over 3D traps, such as their ease of fabrication and integration of trap electronics \cite{Seidelin06, Pearson06}. Consequently, major efforts are underway to fabricate planar ion traps to allow for shuttling of ions through three-way junctions \cite{Amini10,Moehring11}. While many of the building blocks of a planar system have already been demonstrated, transport through planar junctions results in excessive heating \cite{Amini10}.

In ion traps, typically the ions are confined by static fields in one direction and oscillating radio-frequency (rf) fields in the other two directions \cite{Paul90}. If the ion is located at a minimum of these rf fields, it does not feel any forces. Trapping works because when the ion is displaced---e.g. by static electric stray fields---the ponderomotive rf pseudopotential exerts a restoring force on the ion. When the ion is displaced from the field minimum, the rf fields also create a small oscillatory motion of the ion called micromotion. Micromotion can be minimized by choosing static compensation voltages that keep the ion at the rf field minimum. Typically, for ion transport, variable static potentials push the ions along channels formed by constant amplitude radio-frequency (rf) electrodes. However, no planar structure has been found which exhibits such an rf null (place of no micromotion) throughout transport in a Y, T, or X-type junction \cite{Wesenberg09}. As a result, transport through such junctions by constant amplitude rf will always be accompanied by a micromotion of the ion which cannot be compensated.

The resulting micromotion creates a force which repels the ion from the junction, thus making it harder to shuttle the ions adiabatically through the junction. Furthermore, noise close to the secular sidebands of the rf-drive frequency may heat the ions \cite{Blakestad2009}. While transport through an X-junction with low motional heating was recently accomplished in a three dimensional trap structure \cite{Blakestad2011}, transport in a planar trap through X-junctions has yet to be observed. To counter these problems, the electrode structure can be numerically optimized to reduce the micromotion in the junction and transport through Y-junctions has been archieved in planar traps, albeit accompanied with large motional heating \cite{Amini10,Moehring11}. 

Here we investigate the advantages of adjustable amplitude rf for transporting particles through junctions. Transporting via adjustable amplitude rf keeps the ion at the rf null throughout transport, thus avoiding any rf-barrier. This is a novel approach to transporting ions in traps, and is different from current proposals for transporting ions. Some groups have adjusted rf amplitudes to make small displacements in ion position \cite{Cetina07, VanDevender10, Herskind09}. In addition, adjustable rf amplitudes were used to move dust particles close to one another in a planar array of ring traps \cite{Kumph2011}. We propose an alternate scheme for using adjustable amplitude rf as an arbitrarily expandable mode of transport in planar ion traps. Furthermore, our method allows for a straightforward implementation of X-junctions: a notoriously difficult task for conventional transport schemes based on varying static voltages \cite{Blakestad09}, and something not yet achieved in planar traps \cite{Wesenberg09}.

In the first section of this paper, we describe a 3D trap for dust particles demonstrating the basic concept of rf transport. We implement a six-way junction and transport charged particles smoothly in three dimensions between trapping zones by adjusting the rf pseudopotentials. In the second part, we suggest a planar ion trap design for scalable QIP.
Our design includes X-junctions through which ions can move without micromotion. The design is compact because subsequent junctions can be placed close to one another. Keeping ions at the rf null throughout transport allows for easy characterization of the transport procedure spectroscopically and avoids heating due to noise on the secular sidebands of the rf-drive. This allows for reliable transport with little heating. 

\section{Three Dimensional Trap for Dust Particles}

\begin{figure}
\begin{center}
\includegraphics[width = 3 in]{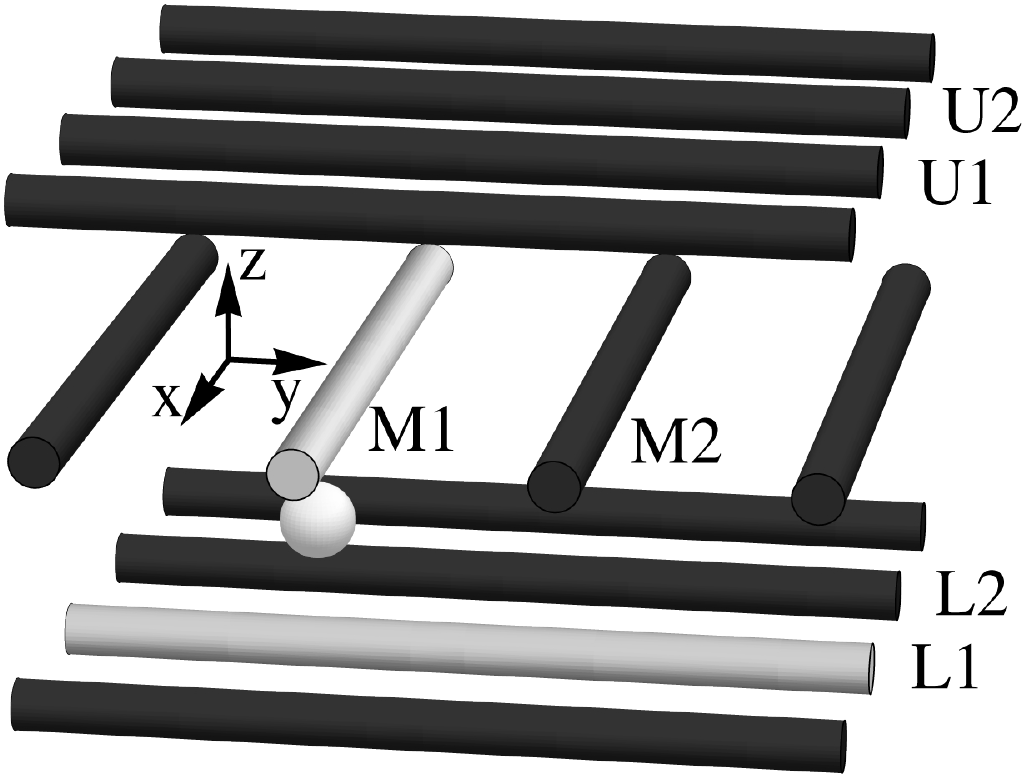}
\caption{\textsc{Trap Dimensions and Design:} The trap consists of three sets of coplanar rods alternating in x and y directions. Note that the trap design could be repeated an arbitrary number of times in all directions. In contrast to what is discussed in Ref.~\cite{Hucul2008}, we apply rf to the grey electrodes and ground the black electrodes, trapping occurs at the intersection where the ball is shown. In the experiments we describe, adjustable rf signals are applied only to the labeled electrodes. L refers to lower, M to middle and U to upper. The distance between the centers of neighboring electrodes and layers of electrodes is 10~mm. Each electrode has a 1~mm radius.}
\label{fig:trapdesign}
\end{center}
\end{figure}

To demonstrate the basic concept of rf transport, we set up a 3D trap for moving charged dust particles in three dimensions using adjustable amplitude rf \footnote{The driving frequency of our trap is 40~Hz. However, we will call these rf electrodes to conform to common ion trapping parlance.} (See Fig.~\ref{fig:trapdesign}). We looked for several characteristics in a model trap: control over location of rf minimum, optical access of trapped particles, scalability to arrays of traps and trap depth. We arrived at the geometry presented already in Ref.~\cite{Hucul2008}. Here however, we apply rf voltage to crossing wires for trapping and employ a different scheme for transporting the particles, as we describe now.

\begin{figure}
\includegraphics{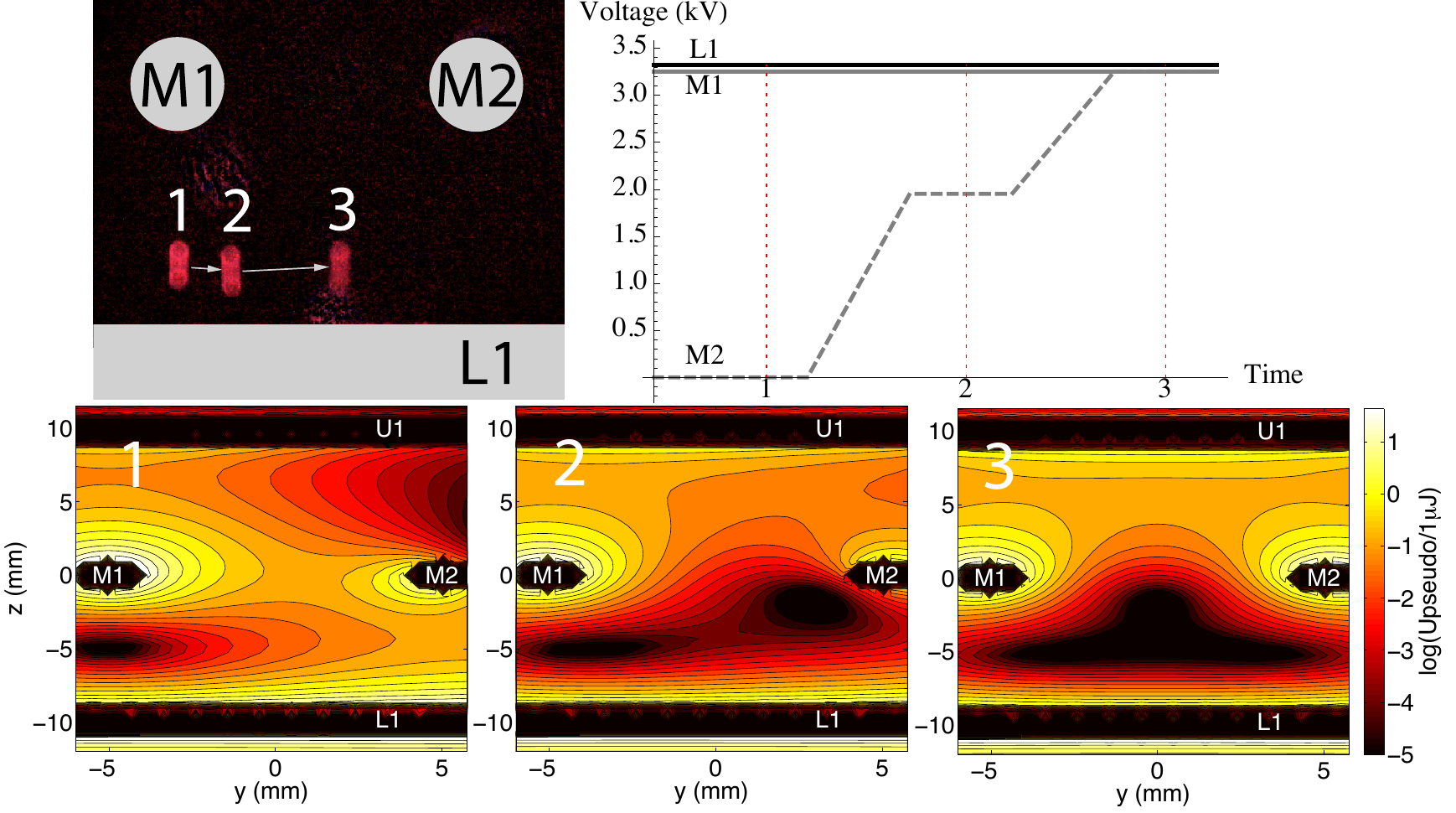}
\caption{\textsc{Horizontal Transport in the 3D trap}: In the top left are overlaid photographs of trapped particles illuminated by a laser. Their positions are labeled by a number and the electrodes are drawn in and labeled for clarity. The voltage ramps corresponding to the particle positions are illustrated on the top right. Below these, we have the modeled pseudopotential contours on the y-z plane, using \(q = 5\times 10^{-9}\) C, \(m = 1 \times 10^{-5}\) kg, \(f = 50\) Hz. The position number is written in white in the top left corner of each contour plot. The contour plots have a lower and an upper cutoff energy which facilitates viewing.
\label{fig:moving_2d}}
\end{figure}

The ion can be transported by changing the electrodes to which rf is applied. Supposing we apply rf to electrodes L1 and M1, trapping occurs between L1 and M1 (as in Fig.~\ref{fig:trapdesign}). The ion can be moved in the positive \(y\) direction by ramping up the rf amplitude on electrode M2. One would then expect the combination of rf on M1 and M2 to act as a single upper electrode; sthe ion will be trapped half way between M1 and M2 above L1. By ramping down the rf amplitude on electrode M1 the ion then completes transport to below M2 and above L1. Fig.~\ref{fig:moving_2d} shows the modeled pseudopotential of this motion scheme, the associated voltage configurations and experimental pictures of the transport. Transport operations in the \(x\) direction are performed in a similar manner by ramping the rf amplitudes on the L electrodes.

We constructed the trap using copper rods with a 1~mm radius as electrodes and Plexiglas plates as a frame. The electrodes are oriented in three sets of planes with wires alternately in the x and y directions (See Fig.~\ref{fig:trapdesign}). The distance between the stacks and between parallel electrodes is 10mm. We load chalk particles into the trap by dipping a statically charged zip-tie (zip-ties appear to be sufficiently charged without any special treatment) into powdered climbing chalk (MgCO$_3$) and flicking the zip-tie against trap electrodes near the trapping site. A Plexiglas box around the trap blocks air currents. Laser pointers illuminate the chalk particles for detection.

We modeled the pseudopotential of this trap using an electrostatics solver \cite{Singer09}. The solver uses the fast multipole method detailed by Nabors \emph{et al.} \cite{Nabors94}. This method works by placing a small patch of charge on an electrode, calculating the potential elsewhere in space using multipole expansions, and adjusting the charge so as to satisfy boundary conditions on
other electrodes. The pseudopotential is calculated using the magnitude of the electric field at
each point:
\begin{equation} \label{eq:Upseudo} \mathrm{U}_{\mathrm{pseudo}} =
\frac{q^2}{4m\Omega^2} E_{0}^2, \end{equation}
where \(q\) and \(m\) are the charge and mass of the ion, \(\Omega\) is the drive frequency on the rf electrodes, and \(E_0\) is the maximum amplitude of the electric field due to the rf electrodes at a given point \cite{Gerlich92}.
Transport in the trap requires electronics that can produce rf signals that are independently adjustable on neighboring channels. A microcontroller (Atmel AtMega32A) controls a DAC (Maxim Max 534),
outputting a 0-5V 40~Hz sinusoid. This signal is electrically isolated using an optocoupler (4N35); a high pass filter removes the resulting offset. The signal is amplified by a high voltage op amp (PA240CC) to a maximum of 100~V. This is further amplified by a transformer (120~V--3.3~kV). We measure the output with a voltage divider.

We measured the charge-to-mass ratio of the trapped particles using the relation between the amplitude of the micromotion \(\xi\) and the electric field strength \(E_0\) (from \cite{Gerlich92})
\begin{equation} \label{eq:micromotion} {\xi} = \frac{q E_0}{m \Omega ^2} \qquad . \end{equation}
We moved the particles to various trap positions using a static potential electrode, measured the amplitude of the micromotion at each position and calculated the theoretical electric field value at each point. The charge-to-mass ratio of a trapped particle was found to be $+5.7 \times10^{-4}$~C/kg with a spread of $\pm0.7 \times10^{-4}$~C/kg.

The main advantage of transport using solely rf fields is minimal micromotion. This is achieved because the particle is always trapped at the rf null, where the electric field strength is zero, provided there are no additional forces. For ion trapping, gravity forces are negligible because of the high charge-to-mass ratio. For trapping dust particles, gravity becomes important because the charge-to-mass ratio of dust particles is ten orders of magnitude lower than that of ions. Thus the particles are pulled out of the rf null, inducing micromotion (this can be seen by the elongated shape of the particle in Fig.~\ref{fig:moving_2d}). The micromotion can be compensated with a static voltage offset that pushes the particle back into the rf null.
On moving horizontally in the \(x\) direction, gravity pulls the particles out of the trap when the particles do not have an electrode below them (in the \(-z\) direction). Thus, compensation of gravity with a static offset to U1 of \(-800 V\) is essential for successful transport. Gravity does not lead to loss of particles moving in the y direction (even without static compensation) because electrode L1 can guide the particles through transport. Once these corrections were made, we completed 1000 transports in \(x\) and \(y\) without loss.

For transport in \(x\) and \(y\), the particle is trapped at the average location of the activated rf electrodes. Thus one would expect that the particle could be transported in \(z\) by starting with electrodes M1, M2 and L1 rf active (Fig.~\ref{fig:moving_2d}, plot 3) and then lowering the rf amplitude of L1. This scheme does not work. As the rf amplitude on L1 is lowered, the trapping zone splits into an upper and a lower zone. When we continue to lower the rf amplitude on L1, the upper zone moves upwards, and the lower zone moves downwards out of the trap. We were able to push the particles into the upper zone via static fields, but we wish to transport using only rf fields.
 
\begin{figure}
\includegraphics{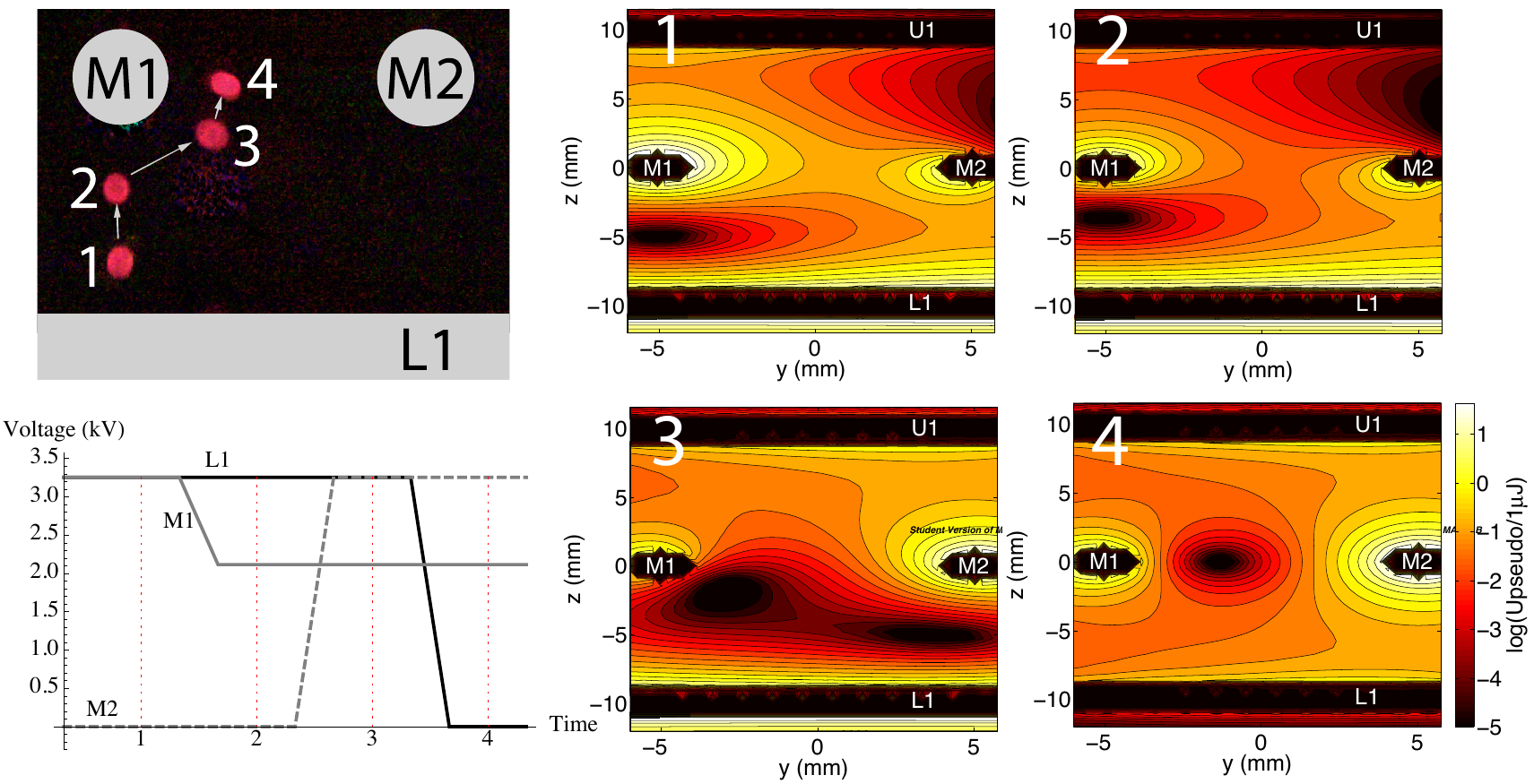}
\caption{\textsc{Vertical Transport in the 3D trap}: In the top left corner, there are four overlaid photographs of trapped particles. Their positions are numbered and the electrodes are drawn in for clarity. In the bottom left corner, the voltages that correspond to particle positions are shown. To the right, there are four modeled pseudopotential contour plots on the y-z plane, corresponding to the particle positions from the photographs. The position number is written in white in the top left corner of each contour plot. Transport from between M1 and M2 into the upper plane (see Fig.~\ref{fig:trapdesign}) is analogous. The contour plots were modeled using \(q = 5\times 10^{-9}\) C, \(m = 1 \times 10^{-5}\) kg, \(f = 50\) Hz.
\label{fig:moving_3d}}
\end{figure}

In order to transport in \(z\) using only rf fields, we applied different rf voltages to M1 and M2 as illustrated in Fig.~\ref{fig:moving_3d}. Starting with the particle at the intersection of M1 and L1, the particle first moves upwards toward M1, and then skirts around electrode M1 towards the center of M1 and M2. This demonstrates transport in \(z\) because the transport is reversible and all other \(z\) transports are the same by symmetry.

The two types of transport demonstrated so far show that transport between arbitrary lattice sites using adjustable amplitude rf is possible. Since the 3D trap is not meant for QIP, but only as a proof of concept for transport via adjustable amplitude rf, we do not optimize the trap voltages to keep trap frequencies constant. Next, we propose a planar ion trap which uses adjustable amplitude rf to facilitate scalable QIP.

\section{Planar Ion Trap Proposal}

Multi-zone planar traps are one of the most promising routes towards scalable quantum information processing with trapped ions \cite{Kielpinski2002}. Here we propose a design that takes advantage of transporting ions using adjustable rf pseudopotentials. Splitting of ion strings can be achieved using static potentials. Our scalable planar ion trap allows for micromotion free transport through all paths of an X-junction.
\begin{figure}
\begin{center}
\includegraphics[width = 6in]{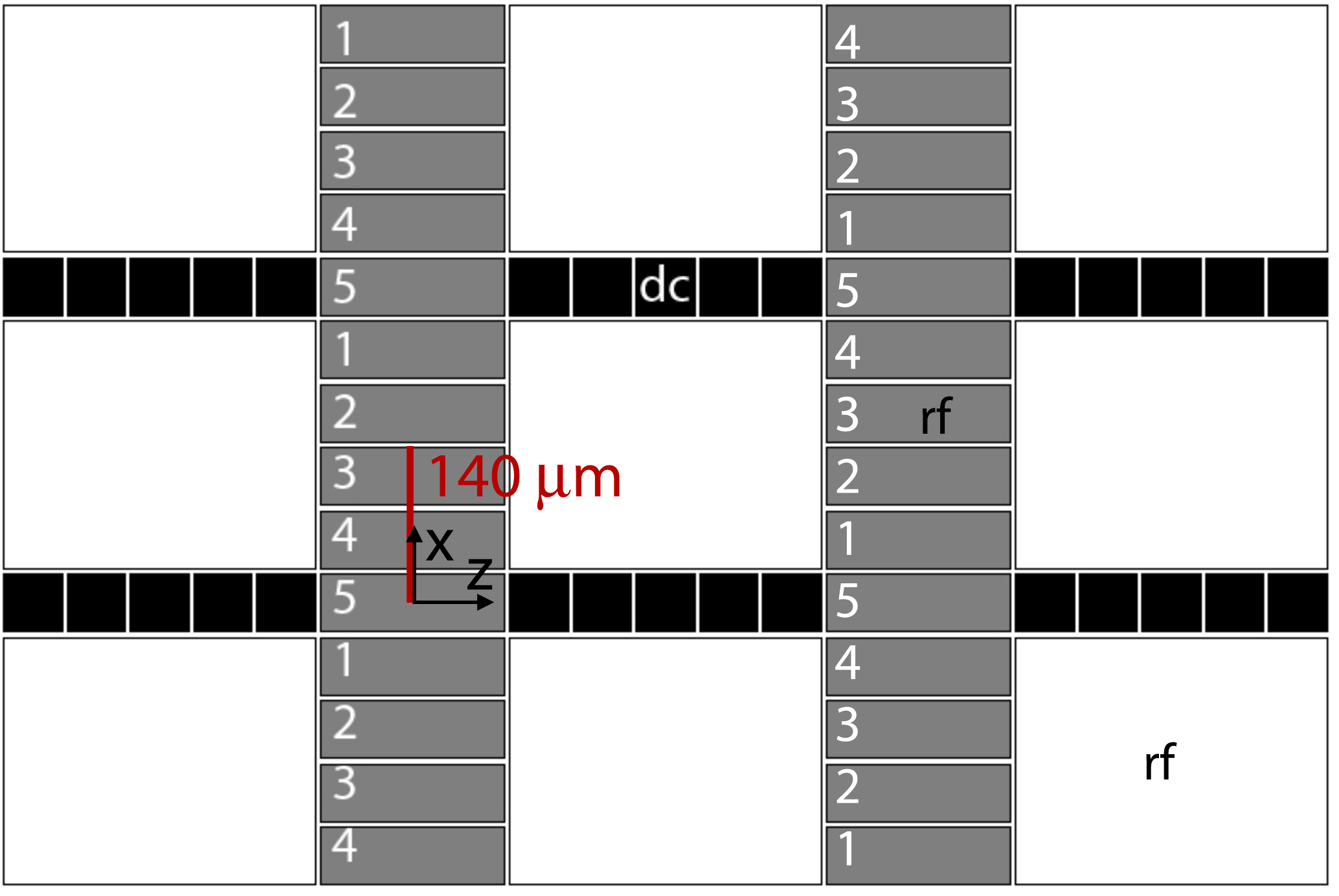}
\caption{\textsc{Planar Trap Design}: The storage and computation channels are
oriented horizontally (black). The large white electrodes are permanent rf electrodes. The rf transport channel (grey) uses adjustable rf signals to transport ions with minimal micromotion. Neighboring rf channels have opposite ordering, thus minimizing their impact on the micromotion in the center of the computation channels. The ion can be moved through the X-junction in any direction. The path (red) from \(x= 0~\mu\)m to 140~\(\mu\)m shows the path on which regularized voltages were calculated (Sec. \ref{sec:optimized-transport}).
\label{fig:planar_trap}}
\end{center}
\end{figure}
The planar design is depicted in Fig.~\ref{fig:planar_trap}. Ions are trapped above long channels of static electrodes (in the \(z\) direction) for storage and computation. The ions can be moved within the computation channel by changing the position of the well created by the static  electrodes. To move ions along the rf channel (the vertical channel), we propose five rf channel electrodes. The central rf channel electrodes are grouped in fours to form an effective rf electrode as wide as the outer permanent rf electrodes.

\begin{figure}
\begin{center}
\includegraphics[width = 6in]{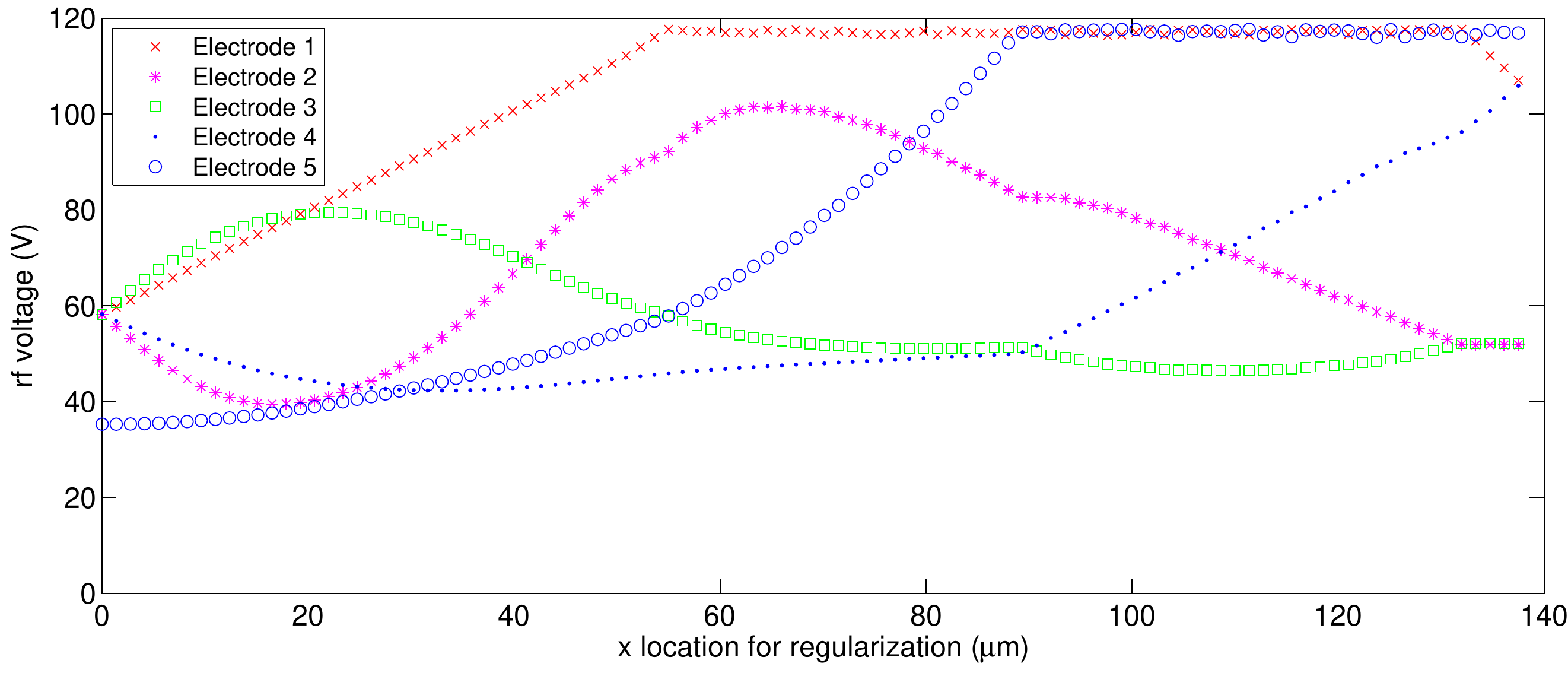}
\caption{\textsc{Regularized trap voltages for rf transport}:
Resulting rf amplitudes from the Tikhonov regularization versus x position with iterative adjustment of weighting factors in order to comply with constraints. The pseudopotential minimum is moved in the positive x direction.
\label{fig:voltages}}
\end{center}
\end{figure}

\begin{figure}
\begin{center}
\includegraphics[width = 4in]{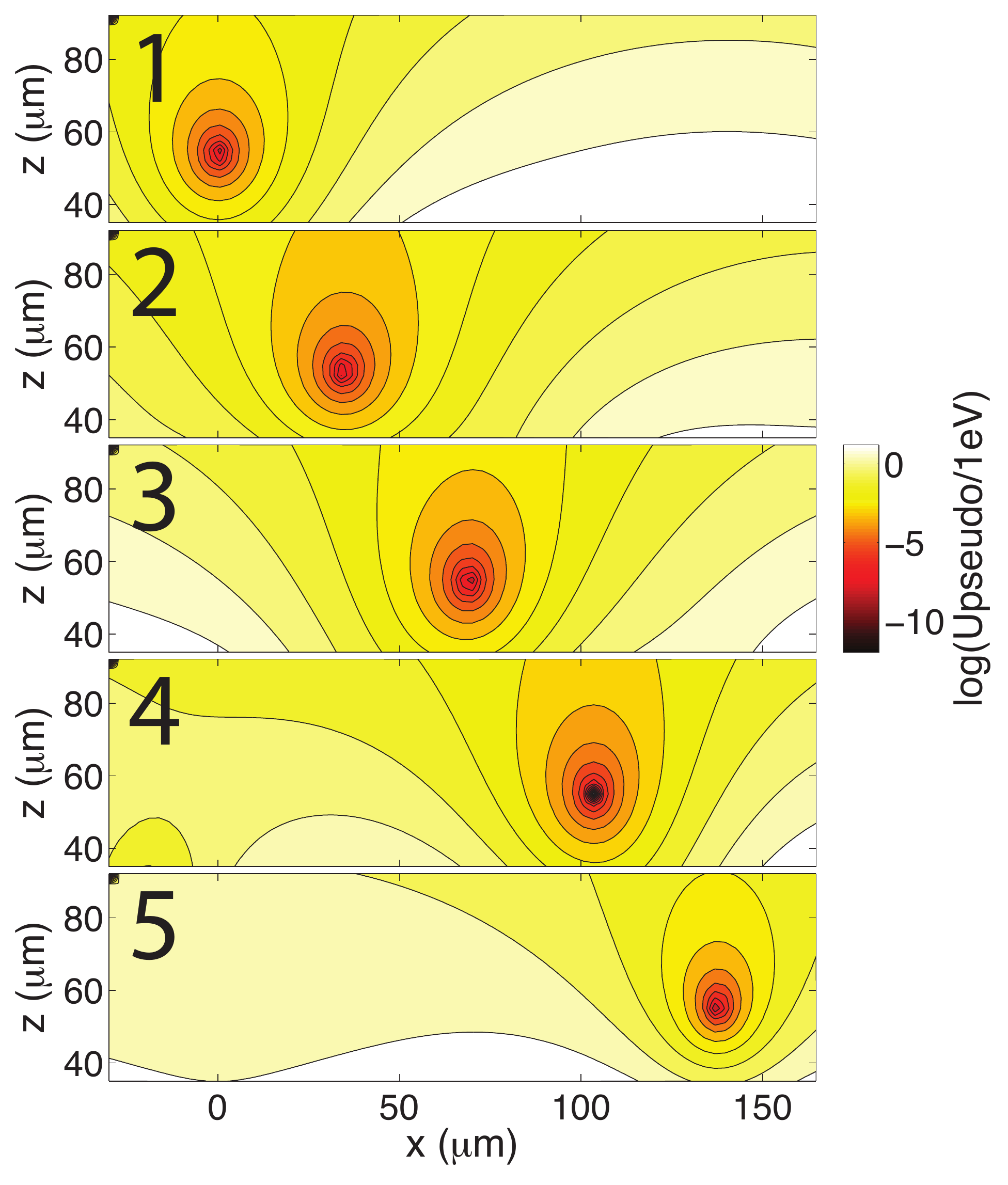}
\caption{\textsc{Pseudopotential contours for planar transport}:
The pseudopotential contours are plotted at 5 points along the transport path. A log plot is used. From top to bottom, the x positions of the 5 points are $0, 34.4, 68.8, 103.1, 137.5 \: \mu m$.
\label{fig:planar_transport}}
\end{center}
\end{figure}

To optimize the trap depth for a given trapping height, the ratio of rf:static electrode widths should be 3.68:1 \cite{Nizamani2010}. Since in our trap rf is applied to four electrodes and the fifth is grounded, we have an rf:static ratio of 4:1. This gives a trap depth of $99.8\%$ of the maximum value \cite{Nizamani2010}. We modify the rf amplitudes of the electrodes (as labeled in Fig.~\ref{fig:planar_trap}) via the scheme shown in Fig.~\ref{fig:voltages}. This motion scheme was found using numerical regularization techniques \cite{Singer09} which keep the trap frequency and depth relatively constant. This is discussed further in Sec.~\ref{sec:optimized-transport}. The particle moves along the transport path depicted in Fig.~\ref{fig:planar_trap}. Fig.~\ref{fig:planar_transport} shows the modeled pseudopotential contours along this transport path. The rf channel acts as a conveyor belt for ions, shuttling all ions that are moved into it in the same direction. Neighboring rf channels shuttle ions in opposite directions, thus suppressing micromotion in the center of the computation channel. The basic design elements---the X junction, computation channel and rf channel---can be repeated many times in a compact way. Local stray-fields would lead to a displacement of the ion from the rf field minimum and can be countered by additional compensation electrodes or by mixing static and rf potentials on the electrodes. In the latter case, resistors are likely to induce phase shifts and proper electronics design and/or phase adjustment of the original drives have to ensure that the rf potentials at the trap have the same phase.

\subsection{Optimized rf transport}
\label{sec:optimized-transport}

We would like to ensure a trapping depth of $25$ meV during the process and avoid parametric heating. Thus we would like to keep the trap frequencies fixed or at least ensure that they change slowly. We also seek to avoid jumps of the trapping position during the transport. To accomplish this we employed the Tikhonov regularization technique presented in \cite{Singer09}. In order to find the rf amplitudes for a desired geometry we used the inverse method on the electric fields generated by the rf electrodes. \footnote{The source code is supplied as supplementary online material and can be downloaded from \texttt{http://kilian-singer.de/ent}.}

A model field was calculated over the junction by keeping electrode 5 at an rf amplitude of 35.3 V and all other rf electrodes at an rf amplitude of 58.2 V, yielding a trap located at a height of 55 $\mu$m over electrode 5. This constitutes the starting condition at \( x = 0 \) $\mu$m. This model field is then translated in the positive x direction sequentially (100 steps) until the minimum is located between electrodes 2 and 3 at \(x = 140~\mu\)m (see Fig.~\ref{fig:planar_trap}). The pseudopotential at the trapping height at five locations along the transport path is displayed in Fig.~\ref{fig:planar_transport}. The other half of the transport is the same by symmetry. The voltages are obtained by the numerical regularization method which calculates the inverse of the field equations. As the large rf electrodes (white in Fig.~\ref{fig:planar_trap}) are kept at an amplitude of 58.2 V, we perform regularization without the field that they create, then add it back afterward. The initial voltages are chosen such that z and y positions of the pseudopotential minimum are kept as constant as possible during transport. If the Tikhonov regularization leads to results exceeding the constraints then weighting factors are iteratively adjusted as described in \cite{Singer09} until the constraints are fulfilled. Whenever the rf amplitude has to be reduced on one electrode, other electrodes are adjusted to compensate while optimizing the trapping potentials.

\begin{figure}
\begin{center}
\includegraphics[width = 6in]{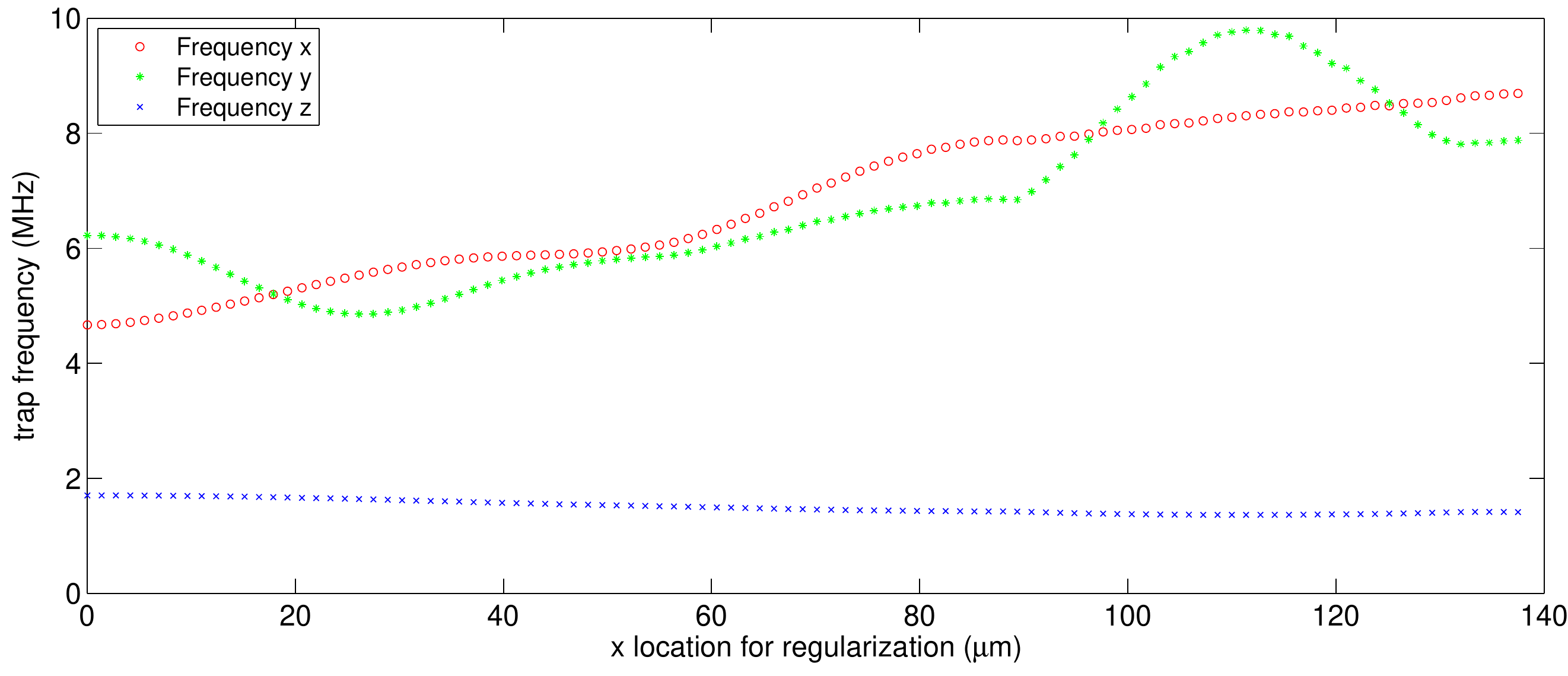}
\caption{\textsc{Trap frequencies}:
Trapping frequencies versus position of the trap minimum as realized by the rf amplitudes from Fig.~\ref{fig:voltages}.
\label{fig:trapfreq}}
\end{center}
\end{figure}

\begin{figure}
\begin{center}
\includegraphics[width = 6in]{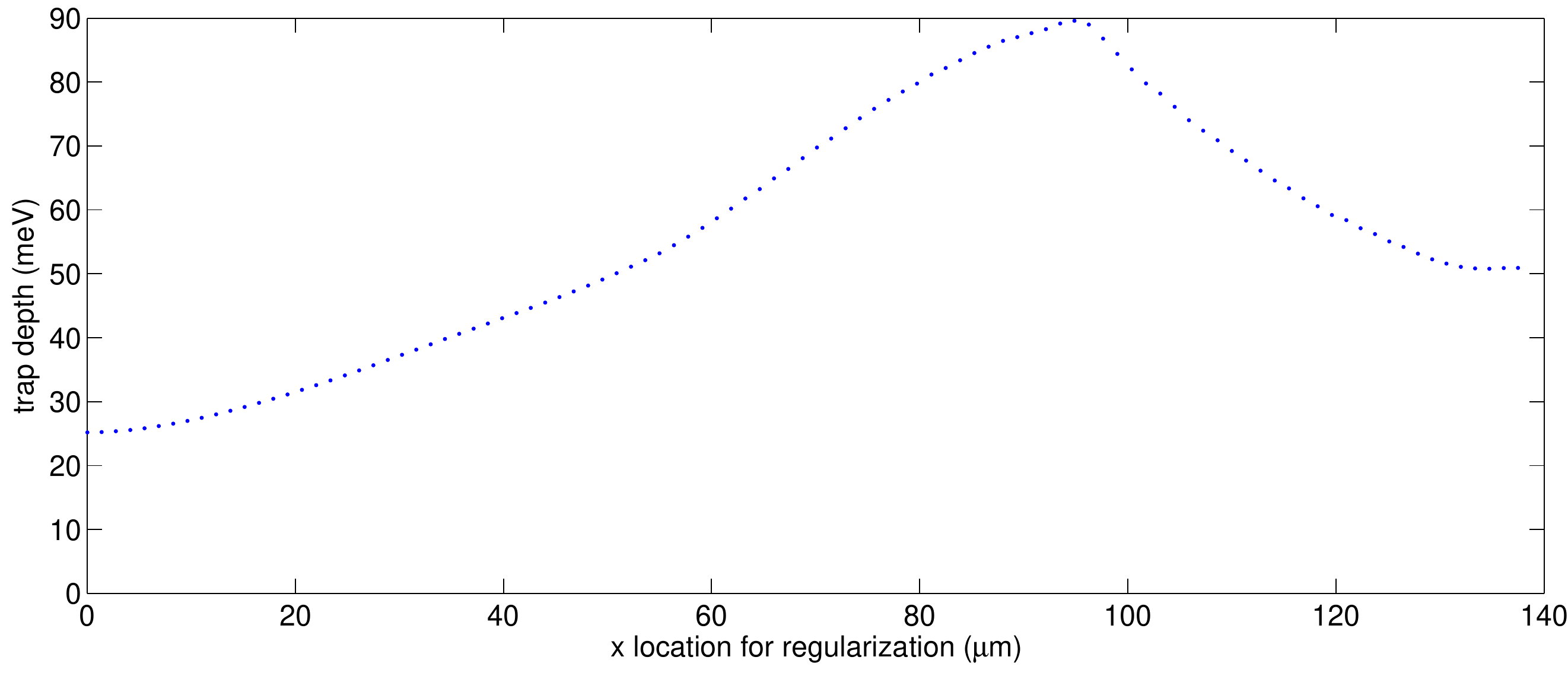}
\caption{\textsc{Trap Depth}:
Minimal pseudopotential trap depth. A minimum trap depth of more than 25 meV can be guaranteed during the whole transport operation.
\label{fig:pseudodepth}}
\end{center}
\end{figure}

The height (\(y\)) of the calculated transport path (Fig.~\ref{fig:voltages}) is 54.4 $\mu$m with a maximum deviation of 1.2 $\mu$m; the z position of the path is -3.9 $\mu$m with a maximum deviation of 1.1 $\mu$m. Smaller deviations from $z=0$ could be achieved if the trap shown in Fig.~\ref{fig:planar_trap} were more symmetrical about the line $z=0$. Fig.~\ref{fig:voltages} depicts the voltages calculated as a result of the Tikhonov regularization. Fig.~\ref{fig:trapfreq} shows the trapping frequencies along all three axes. We expect that the frequency variations can be strongly reduced by relieving some of our stringent boundary conditions: such as allowing for higher maximum voltages, relaxing the  crosstalk constraint to a larger range (currently 20-100V, see Sec.~\ref{sec:rf_design}), and allowing additional static offset of the rf electrodes. We estimate the trap depth by calculating the potential energy along 27 lines through the minimum of the trap and find that the energy barrier is always larger than 25 meV along any of these lines (Fig.~\ref{fig:pseudodepth}).

\subsection{Spatial resolution of rf potentials}

An interesting feature of the rf pseudopotential is that it does not obey the Laplace equation. Thus recognizing general properties of the pseudopotential can be fruitful in application of rf electrodes beyond three-dimensional trapping and shuttling of charged particles in rf traps. Here we show that rf pseudopotentials can be more sharply peaked than static potentials, and thus have a higher spatial resolving power. This can be useful, for example, for selectively exciting segments of ion strings \cite{Pruttivarasin2011}.

\begin{figure}
\begin{center}
\includegraphics[width = 4.6in]{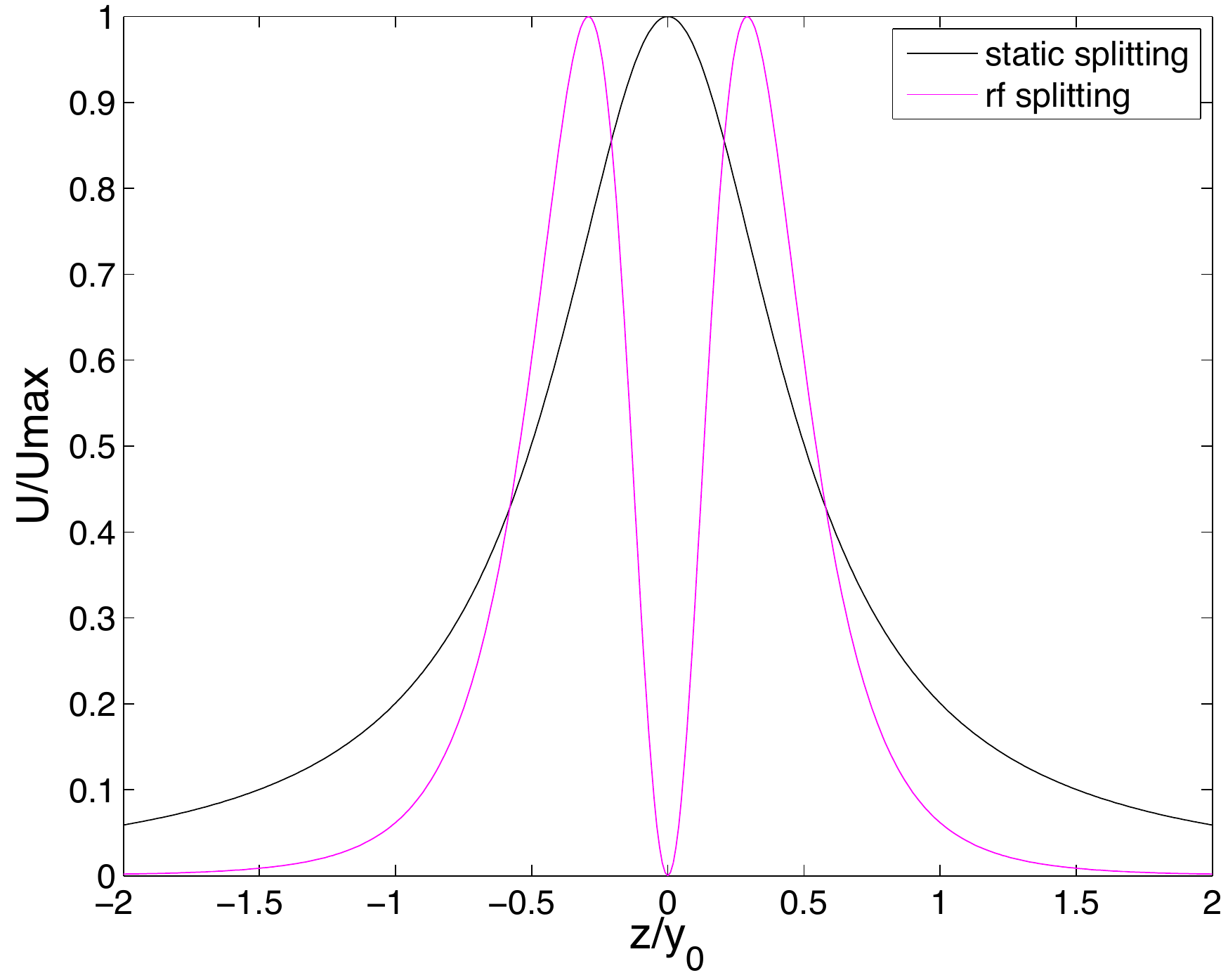}
\caption{Normalized electrostatic potential (black) and rf pseudopotential (magenta) of an axially thin electrode. Potential ($U/U_{\rm max}$) is plotted versus normalized axial position ($z/y_0$) at trapping height $y_0$. Here the axial extent of the electrode is $0.1\,y_0$. The resolving power of rf is a factor of 6 higher than that of a static potential. \label{fig:normalized_potentials}}
\includegraphics[width = 4.6in]{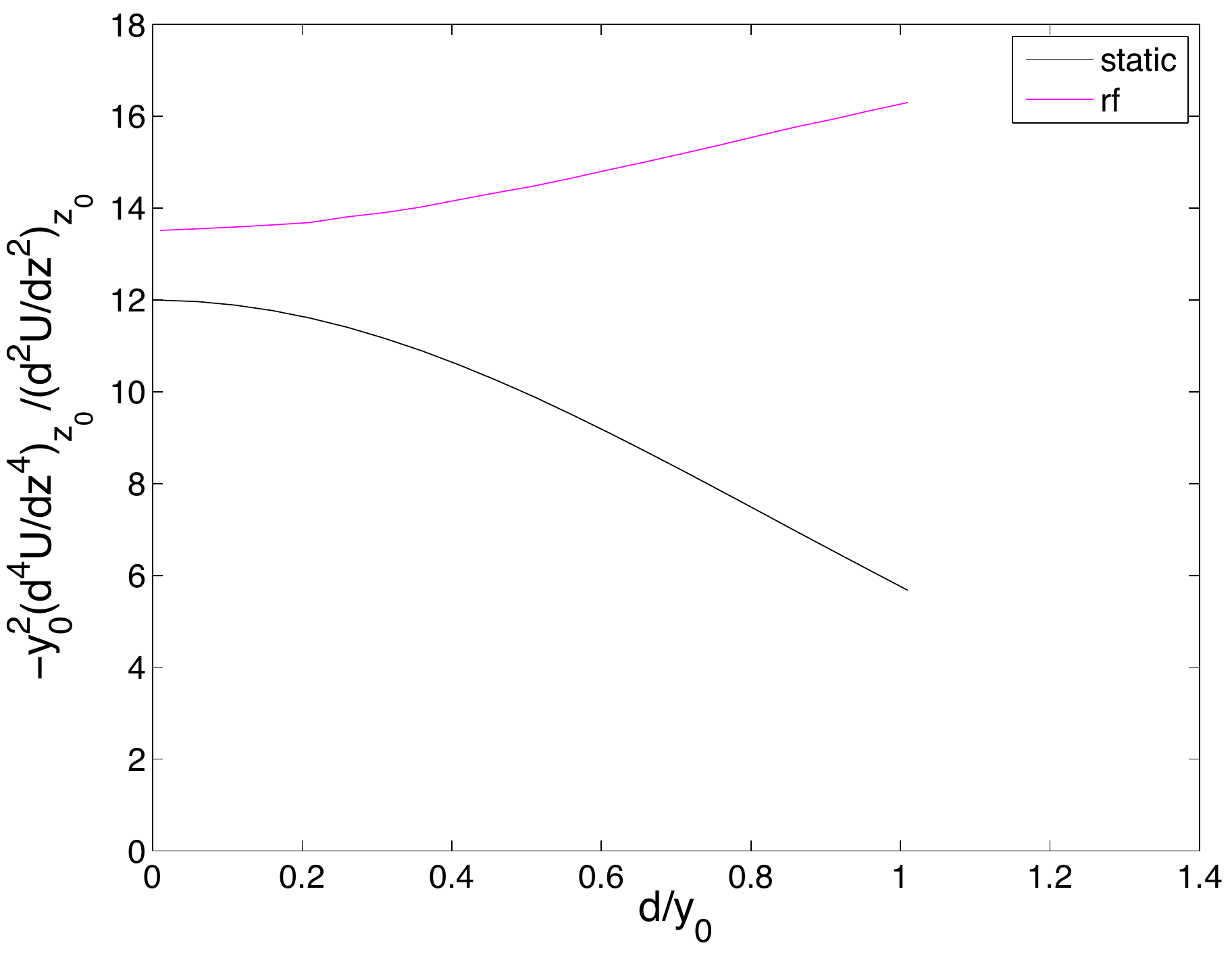}
\caption{Ratio of quartic to quadratic term of static (black) and rf pseudopotential (magenta) for a splitting electrode with axial length d and large lateral width. $y_0$ is the ion height. This ratio determines the minimum center-of-mass mode frequency, see text.
\label{fig:quartic_quadratic_term}}
\end{center}
\end{figure}

Let $U(z)$ be the (pseudo)potential of a single rf or static electrode of a surface trap, such as the one shown in Fig.~\ref{fig:planar_trap}, along the axial direction and at the ion height. This is superposed to the potential of the remaining electrodes, assumed to be held at fixed values. To quantify how ``sharp''  $U(z)$ is, we define the resolving power $r \equiv \left(d^2U/dz^2\right)_{z_0}/U(z_0)$  of such an electrode, by using the normalized second derivative of $U(z)$, at the axial position, $z_0$. Physically, this corresponds to the ratio of the spatial gradient of the force on nearby ions to the disturbance of the overall trap potential due to the single electrode. In order to use an electrode to displace ions in the axial direction, we consider an axially narrow and laterally long rectangular electrode (i.e. having high-aspect ratio $x_{\rm e}/z_{\rm e}$, where $x_{\rm e}$  and $z_{\rm e}$ refer to the x and z extents of the electrode respectively). For such an electrode, the potential evaluated at the ion height has a position of highest curvature directly above the middle of the electrode for static potentials (see Fig.~\ref{fig:normalized_potentials}(a)). In comparison, for rf electrodes there are two positions of highest curvature, axially displaced from the electrode by approximately $\pm \sqrt{3} y_0$, where $y_0$ is the ion height. For a trapping height of 100~$\mu$m and an electrode with an axial width $z_{\rm e} = 10\:\mu$m, we numerically determine the resolving power of the rf pseudopotential to be a factor of 6 higher than that of a static potential. In the limit of an rf electrode with vanishingly small extent in the axial ($z$) direction, we find that the resolving power of the rf pseudopotential is 4.5 times larger than that of a static potential.

In the context of ion shuttling, it is tempting to use rf electrodes for splitting of ion strings, and we investigated this possibility. We considered simple geometries involving only one splitting electrode, used to modify a harmonic trapping potential. During the splitting of a two-ion crystal, the minimum center-of-mass mode frequency for a given harmonic potential is determined by the ratio of quartic to quadratic coefficients of the axial (pseudo)potential at the maximum of the (pseudo)potential,  $z_0$ , where the string splitting occurs \cite{Home2006a}. This ratio is plotted in Fig.~\ref{fig:quartic_quadratic_term} for splitting electrodes of different axial extents. Despite the significantly higher ratios for the pseudopotential compared to a static potential, the increase in minimum center-of-mass mode frequency is less pronounced due to the $\omega_{\rm min} \sim \left((d^4U/dz^4)/(d^2U/dz^2)\right)^{3/10}$ dependence \cite{Home2006a}. Moreover, a technologically appropriate figure of merit for splitting of ion strings is the absolute strength of the achievable octupole term in the potential for a given maximal voltage \cite{Home2006a}. Therefore, the increased resolving power of rf potentials does not necessarily translate to an improved ion string splitting performance. Nevertheless, it cannot be ruled out that the use of pseudopotentials to augment static potentials in ion-string splitting can be useful and a numerical investigation of particular trap geometries may be worth carrying out.

\section{RF design considerations}\label{sec:rf_design}
Creating several adjustable high voltage rf channels in an ion trap could be a challenge due to capacitive coupling between different electrodes. Suggestions for the implementation of rf control include changing tapping location of a helical resonator continuously \cite{VanDevender10} and varying load capacitances in a resonating circuit \cite{Herskind09}. Fast switching of rf amplitudes can be achieved by using multiple step-up resonators, where each resonator is used to drive only one rf electrode. Helical resonators are hard to shield and as a result the various resonators connected to different rf electrodes will be strongly coupled to each other. Therefore, we propose using multiple coaxial resonators (consisting of ordinary coaxial cables in a T-junction) terminating on the different trap electrodes, thus shielding the resonators from each other very effectively.

We tested the degree of control over the rf amplitude with coaxial resonators by applying 100~V at 50~MHz to neighboring electrodes of a test trap (100~\(\mu\)m \(\times\) 2000~\(\mu\)m electrodes, separated by 10~$\mu$m) \cite{Daniilidis09}. The trap was mounted on a chip carrier which resulted in capacitive coupling of approximately 8~pF between neighboring electrodes, and a resistance to the electrodes of approximately $0.4~\Omega$, determined by the wire bonds between the trap and chip carrier. We added a capacitance of 10~pF from each rf electrode to ground to minimize the pickup from the neighboring electrode. The high voltage rf was generated with an rf signal generator, a 5~W rf amplifier and a coaxial resonator consisting of an ordinary RG58 coaxial cable for step up. With this set-up we successfully varied one channel from 20--100~V amplitude while holding the neighboring electrode at 100~V amplitude. The lower rf voltage limit of 20~V on the trap electrodes arises from residual coupling between resonators. An additional complication of the residual coupling between different rf channels were significant phase shifts at low voltages, up to $\pi/2$, during the amplitude variation. Such phase shifts will induce micromotion and are undesirable. These phase shifts can be significantly improved by a better rf design and by using actively phase-locked rf drive circuits \cite{Kumph2011}.

Since the pseudopotential scales with the square of the electric field amplitude, reducing the rf amplitude on an electrode to 20\% is equivalent to reducing the force due to the pseudopotential to 4\%. Our simulations show that transport is hardly affected by a pick-up of this size. In Sec.~\ref{sec:optimized-transport}, we showed how a micromotion free transport is still possible with this constraint.

As already mentioned, one point of concern with various independent radio frequency sources is that a phase difference between different electrodes will result in ion micromotion. A conservative estimate using the trap geometry and parameters in Sec.~\ref{sec:optimized-transport} results in the requirement that the phase difference between different rf electrodes be less than 0.5~mrad in order to maintain micromotion amplitude lower than 10~nm. If this tolerance cannot be met by design, it is always possible to use active phase-locking schemes for the rf drive and to calibrate and adjust the phase of the independent rf-drives. While such technology still needs to be developed for ion traps, we anticipate no fundamental obstacles to its implementation.

In transport schemes using only static potentials, the factors limiting speed are currently the DACs, amplifiers, and filters used for the static electrodes. In case of rf transport, an additional limiting timescale is set by the quality factor of the resonant circuit used to generate rf voltages, $\tau=Q/\Omega$ as it will limit the rate with which the amplitude of the rf electrodes can be changed. For a typical Q factor on the order of 100, $\tau = 0.3\:\mu$s  ($\Omega = 2\pi \times 50$~MHz) and is thus not limiting adiabatic transporting schemes. An additional concern regarding the motional state of the ion during transport arises when the secular frequencies become degenerate, as energy exchange between different modes can then occur. In cases where frequency crossings cannot be avoided, either sufficient cooling of all modes before the transport or cooling after the transport can mitigate this problem.

\section{Conclusion}

We propose micromotion-free transport of trapped ions by changing the amplitude of rf electrodes in Paul traps. We demonstrate a smooth transport of charged dust particles between arbitrary lattice sites in three dimensions. In principle, our method allows nearly arbitrary positioning of particles in 3D restricted only by trap depth considerations and the space the electrodes take up.

Based on these results, we propose a planar architecture for scalable QIP using X-junctions which are not achievable with conventional transport schemes. Furthermore, particles can be transported without micromotion, reducing potential ion heating during transport and allowing for easy characterization of the transport path. We present optimized amplitude ramps using specially adapted regularization techniques. Our compact design allows for high fidelity X-junctions and thus presents a new tool to scale ion trap quantum computers. \\
\\
The experiments were supported by the Laboratory Directed Research and Development Program of Lawrence Berkeley National Laboratory under U.S. Department of Energy Contract No. DE-AC02-05CH11231. KS acknowledges additional financial support by the VolkswagenStiftung. ND was supported by the Office of the Director of National Intelligence (ODNI), Intelligence Advanced Research Projects Activity (IARPA), through the Army Research Office. We also acknowledge J. Home for bringing reference \cite{Home2006a} to our attention and Tobias Sch\"atz and the referees for helpful comments on the manuscript. \\
\bibliographystyle{nature}
\bibliography{bibliography}
\end{document}